\documentclass[10pt]{article}

\makeatletter
\let\@fnsymbol\@arabic
\makeatother

\usepackage{multicol,caption}
\usepackage{sidecap}
\usepackage{graphicx}
\usepackage{fullpage}
\usepackage[superscript]{cite}

\newcommand{\footremember}[2]{\footnote{#2}\newcounter{#1}\setcounter{#1}{\value{footnote}}}
\newcommand{\footrecall}[1]{\footnotemark[\value{#1}]} 

\newenvironment{Figure}
  {\par\medskip\noindent\minipage{\linewidth}}
  {\endminipage\par\medskip}

\begin{document}

\title{Searching for the rules that govern hadron construction}

\author{
Matthew R. Shepherd\footremember{iu}{Indiana University, Bloomington, IN  47405, USA}, 
Jozef J. Dudek\footremember{odu}{Old Dominion University, Norfolk, VA 23529, USA}$^{~,}$\footremember{jlab}{Jefferson Lab, Newport News, VA  23606, USA},
Ryan E. Mitchell\footrecall{iu}
}

\date{March 28, 2016}

\maketitle

\subsection*{Preface}


Just as Quantum Electrodynamics describes how electrons are
bound in atoms by the electromagnetic force, mediated by exchange of
photons, Quantum Chromodynamics (QCD) describes how quarks are bound inside
hadrons by the strong force, mediated by exchange of gluons.
At face value, QCD allows hadrons constructed from increasingly many quarks
to exist, just as atoms with increasing numbers of electrons exist,
yet such complex constructions seemed, until recently, to not be present 
in nature.  In what follows we describe advances in the
spectroscopy of mesons that are refining our understanding of the
rules for building hadrons from QCD.

\vspace{0.6in}

\begin{multicols}{2}

\section*{Introduction}


While decades of experimental study support QCD as the underlying
theory of quark interactions, a detailed understanding of the way QCD
generates protons, neutrons, and other strongly interacting ``hadrons"
remains elusive. The majority of observed hadrons fall neatly into only 
two very limited sets:  baryons consistent with being three-quark 
constructions ($qqq$), and mesons as
quark-antiquark ($q\bar{q}$). As well as constructions featuring 
larger numbers of quarks, QCD also appears to allow hadrons 
built not only from quarks, but also from gluons. 
This has raised the question of why, until possibly now, has there been no evidence 
for a spectrum of such hadrons.  Have we just been historically 
unsuccessful in producing these exotic particles in the laboratory, or
are there more restrictive rules for building hadrons that are not obvious
from the unsolved equations of QCD?  Here we choose to focus specifically
on the spectrum of mesons where timely developments in both
theory and experiment can be used to illustrate how the field of hadron
spectroscopy addresses fundamental questions about QCD, questions
that are common to both the meson and baryon sectors.

\subsection*{Interacting quarks and gluons in QCD}

Within QCD, the ``charge" that controls the interactions of quarks is known
as ``color", and it was study of the empirical spectrum of hadrons that
first introduced the concept of quarks and their threefold color charge.
Interactions in QCD are symmetric under changes of color, {\it i.e.}, no single
color of quark behaves differently from the other two, and imposing this
symmetry on the theory uniquely defines the interactions allowed in QCD
between the quarks and the force-carrying gluons. Colored quarks can 
interact by emitting or absorbing gluons, and because they carry color 
charge themselves, gluons can also emit and absorb gluons.

While observations about the spectrum of hadrons inspired the fundamental
theory of quark interactions, calculating the detailed spectrum from this
theory has so far been impossible. The difficulties in these calculations
stem from the presence of gluon-gluon interactions, which make QCD
interactions very strong on the $10^{-15}$~m distance scale that
characterizes hadrons.  This ultimately results in a property
called ``confinement," whereby quarks are permanently trapped inside
composite hadrons, making it difficult to isolate the interaction of a single
quark and antiquark from the collective behavior of quarks and gluons in
the hadron.  The strong coupling means that, unlike the electromagnetic force, where exchange 
of two photons between electrons in an atom is far
less probable than exchange of just one, exchange of any number of gluons
between quarks in a hadron is every bit as probable as exchanging one.  Because of
this, there is no simple method to calculate the net effect of interactions
between two quarks, and a QCD calculation of the mass of a hadron, easily
measurable by experiment, becomes intractable.

\subsection*{Understanding QCD via rules for building hadrons}

Our inability to solve the equations of QCD is not just a curiosity -- it
restricts our understanding of the behavior and structure
of hadrons, owing to the lack of any simple relationship between the
fundamental quarks and gluons of QCD and the spectrum of hadrons 
observed experimentally. This has motivated the use of heuristic models, or ``rules", that serve as a
bridge between QCD and experiment, capturing the important features of the
spectrum while attempting to respect the known properties of
QCD. It is the development of a rulebook for construction of hadrons
that is consistent with both QCD and experimental data that arguably
defines what it means to understand how QCD generates hadrons.  A 
uniform set of rules may not exist -- there may be no simple way to capture the
complex behavior of QCD -- but the high degree of regularity in the
experimental spectrum of hadrons suggests that this is not a forlorn hope,
and the search for this rulebook drives the field of hadron
spectroscopy.

An important area of exploration attempts to create previously
unobserved classes of hadron in the laboratory, such as quark-gluon hybrids
or tetraquarks. From the pattern of such states, or their absence, 
we can refine our understanding of the rules of hadron construction. A second area develops techniques for
calculating the observable properties of hadrons directly from QCD, which 
will indicate how the rules follow from the strong interactions of quarks and gluons prescribed by that theory.

In what follows we will review exciting contemporary developments 
in each of these areas and discuss future prospects for achieving the goal
of determining the rulebook for hadron construction.

\section*{Rules inferred from experimental data}

We label hadrons by their mass and their quantum numbers $J$ (spin), $P$
(parity, behavior under reflection in a mirror), and $C$ (charge-conjugation, 
behavior under exchange of particles with antiparticles).  These
properties are directly observable, but other characteristics, such as their
internal composition, must be inferred.  As the number of observed hadrons
has increased over the last half-century, definite patterns have emerged that 
have led to an initial set of simple rules for the construction of hadrons from quarks.

\subsection*{The quark-antiquark rule for constructing mesons}

\begin{figure*}
\centering
\includegraphics[width=0.83\linewidth]{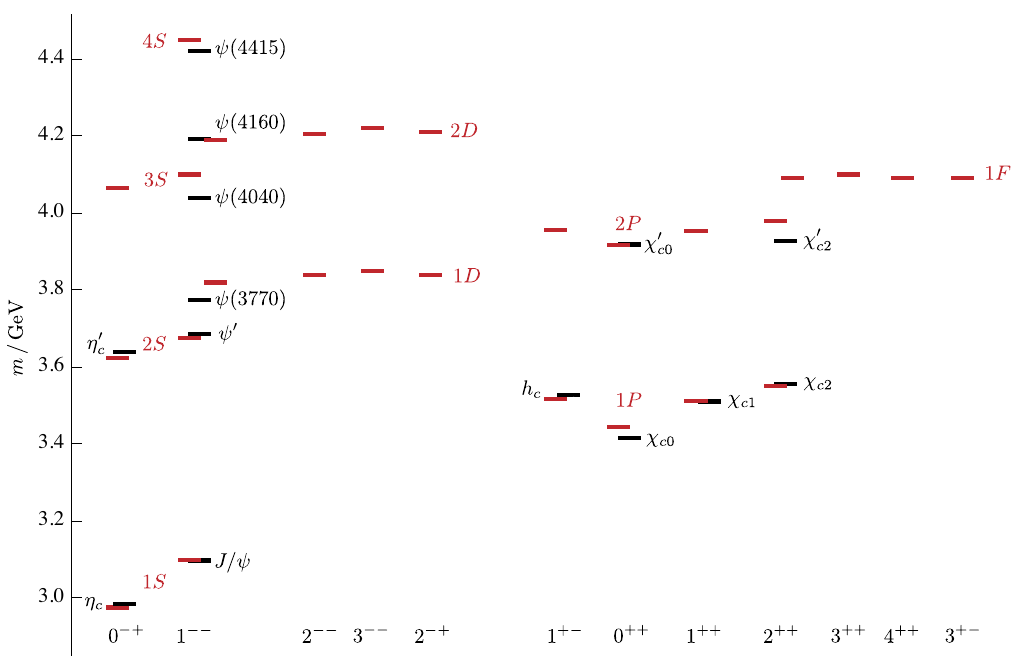}
\caption{\label{fig:charmonium} A $q\bar{q}$ potential model calculation (red)~\cite{Godfrey:1985xj} of the charmonium
spectrum in comparison to experiment (black)~\cite{Agashe:2014kda}. Columns indicate states of common $J^{PC}$. Potential model states appear in groups labelled by their radial and orbital angular momentum quantum numbers, $nL$.} 
\end{figure*}

One of the earliest patterns discovered (in the 1960's) was that mesons
with the same $J^{PC}$ quantum numbers could be grouped into sets of nine
having similar mass. This could be explained by combining a quark $q$ with
an antiquark $\bar{q}$ if there were three ``flavors" of quark -- these
were given the names up, down and strange. The lightest such ``nonet" of
mesons has $J^{PC} = 0^{-+}$, and there are heavier nonets with other $J^{PC}$. It was suggested that
the additional mass-energy of the excited hadrons arises principally from the orbital or radial motion of the
$q\bar{q}$ pair, in analogy to the excitations of a single-electron atom.

With the discovery of charmonium (in the 1970's)~\cite{Aubert:1974js,Augustin:1974xw}, this
quantum-mechanical picture became more precise -- these new mesons with
masses much larger than those observed earlier were explained as being
bound states featuring a new, heavier quark, which was dubbed ``charm". Charmonium
mesons with a range of $J^{PC}$ were observed and their spectrum (Fig.~\ref{fig:charmonium})
resembles that of a pair of particles bound by a potential. The large mass of the charm 
quark justified such an approach, as many of the complexities of a relativistic 
system could be neglected. The potential
needed to describe the spectrum was novel, featuring a steadily rising
piece at large distances that would confine the quarks within the meson~\cite{Eichten:1978tg}. A
feature of this model of mesons is that it is not possible for a
quark-antiquark pair in any orbitally or radially excited state to have $J^{PC}$
in the set $0^{+-}$, $1^{-+}$, $2^{+-}$, \ldots.  Sets of mesons with these ``exotic" quantum
numbers were not convincingly observed experimentally, either
in charmonium or for the lighter quarks, supporting the $q\bar{q}$ picture.

Until recently virtually all experimentally observed hadrons could have 
their presence explained by a simple rule stating that each meson is 
constructed from a quark-antiquark pair, and
each baryon from a three-quark configuration. However it has never been 
at all obvious why QCD chooses to be so parsimonious -- why are there not 
meson-like states of two quarks and
two antiquarks, ``tetraquarks", or baryon-like states of four
quarks and an antiquark, ``pentaquarks"? Furthermore, since the gluons of
QCD strongly interact just as quarks do, couldn't we have ``hybrid mesons" in
which gluons bind to a $q\bar{q}$ pair, and ``glueballs" which do 
not require quarks at all? Observation of hadrons like these would challenge 
the simple rule outlined above, and indeed, recent experimental results 
are casting doubt on how parsimonious QCD really is.

\subsection*{Recent results challenge the quark-antiquark rule}

\begin{SCfigure*}
\centering
\includegraphics[width=1.2\linewidth]{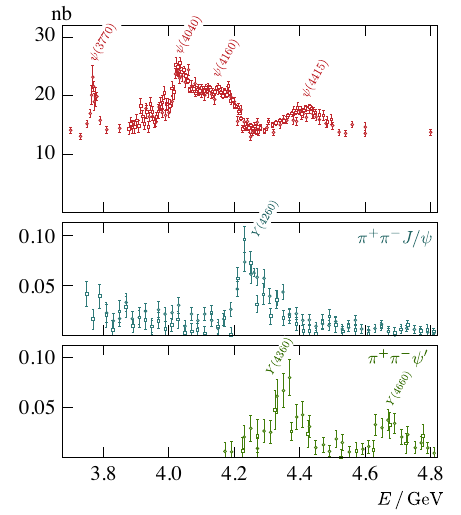}
\caption{\label{fig:epem_hadrons}The cross sections for (top to bottom) $e^+e^- \to \mathrm{hadrons}$~\cite{Bai:2001ct,Osterheld:1986hw}, $e^+ e^- \to \pi^+ \pi^- J/\psi$~\cite{Lees:2012cn,Liu:2013dau,Ablikim:2013mio}, and $e^+ e^- \to \pi^+ \pi^- \psi'$~\cite{Lees:2012pv,Wang:2007ea}. 
The $1^{--}$ states $\psi(3770)$, $\psi(4040)$, $\psi(4160)$ and $\psi(4415)$, indicated in the upper panel, can be associated with the $1D$, $3S$, $2D$ and $4S$ states of the potential model of Figure~\ref{fig:charmonium}. The enhancements observed in the lower panels do not line up with these states, which may indicate that they correspond to new hadron states that do not appear in the potential model and hence do not obey the $q\bar{q}$ rule.} 
\end{SCfigure*}

A powerful way to study the meson spectrum is to collide high-energy beams
of electrons and positrons and to observe the rate at which systems of
hadrons are produced.  In this process, the $e^+e^-$ pair first annihilates, producing a photon; 
the photon converts into a quark and antiquark, which then 
interact, exchanging gluons and perhaps creating more
quark-antiquark pairs; finally, these quarks and gluons arrange themselves into a 
system of hadrons that are observed by the particle detector. If
the collision energy is close to the mass of a meson with $J^{PC}=1^{--}$ quantum numbers, the system
``resonates", and the probability of a collision increases. Thus, a plot
of the normalized rate of hadron production, the ``cross section", against $e^{+}e^{-}$ center of mass energy,
shows bumps corresponding to the produced meson states, also known as ``resonances" (Fig.~\ref{fig:epem_hadrons}(top)).
These excited states live only briefly before decaying into the set of observed lighter hadrons, 
and the width of the bump is inversely related to the lifetime of the state. 

Figure~\ref{fig:epem_hadrons}(top) depicts the {\em total} rate of hadron production
as a function of $e^{+}e^{-}$ center of mass energy.  The bumps are
interpreted as evidence for a series of excited states -- the $\psi(3770)$,
$\psi(4040)$, $\psi(4160)$, and $\psi(4415)$ -- consistent with expectations from the
quark-antiquark picture (see Fig.~\ref{fig:charmonium}). But recent experimental advances
have allowed a closer inspection. If instead of the total rate, we look at the
rates for the production of specific systems of hadrons, distinct features appear 
that have no simple explanation in the $q\bar{q}$ picture.

The production rate of the $\pi\pi J/\psi$ system\footnote{The $J/\psi$ is one hadron 
which for historical reasons has two names associated with it.}, shown in Figure~\ref{fig:epem_hadrons}(middle), provides one such example.  
Here, a prominent bump appears at 4260~MeV, which, surprisingly, lies between the
masses of the $\psi(4160)$ and $\psi(4415)$ states.
Unlike the $\psi(4160)$ and $\psi(4415)$, this ``$Y(4260)$"
resonance has no explanation within the $q\bar{q}$ picture. 
Another example is the production rate of the $\pi\pi\psi'$ system.
The $Y(4260)$ might be expected to also appear here, since $\pi\pi J/\psi$ and $\pi\pi\psi'$ are 
very similar systems, but it does not.  Instead, two
bumps appear, a ``$Y(4360)$" and a ``$Y(4660)$" (Fig.~\ref{fig:epem_hadrons}(bottom)), in further
disagreement with the spectrum suggested by the total cross section. These
$Y$ states, which appear in addition to those expected within the $q\bar{q}$
picture, may be a signal that QCD does indeed produce mesons with internal structure beyond the simple $q\bar{q}$ ``rule".

The observation of new states in charmonium, which was previously believed
to be rather well understood, has spurred a program of searches for further
exotic candidate states, observations of which are providing still more challenges for the simple $q\bar{q}$ rule.
For example, a detailed study of the $\pi^+\pi^- J/\psi$
system produced in $Y(4260)$ decays showed that the $\pi^\pm J/\psi$ system
(Fig.~\ref{fig:zc}(top)) appears to resonate at a mass of 3900~MeV, producing an
electrically charged state labelled $Z(3900)$~\cite{Ablikim:2013mio,Liu:2013dau,Xiao:2013iha}.  This structure is particularly
noteworthy because its large mass and decay featuring $J/\psi$ suggest that it contains
a charm quark and an anti-charm quark, while its net electric charge requires
further light (up and down flavored) quarks. It is thus a candidate for a tetraquark.  A
pattern of such states is beginning to emerge around 4~GeV: e.g. in the $\pi^+\pi^- h_c$ system,
also produced in $e^+e^-$ collisions, another
electrically charged structure, the $Z(4020)$, appears in the $\pi^\pm h_c$ spectrum~\cite{Ablikim:2013wzq}
(Fig.~\ref{fig:zc}(bottom)) with a somewhat larger mass.

\begin{Figure}
\centering
\includegraphics[width=0.9\linewidth]{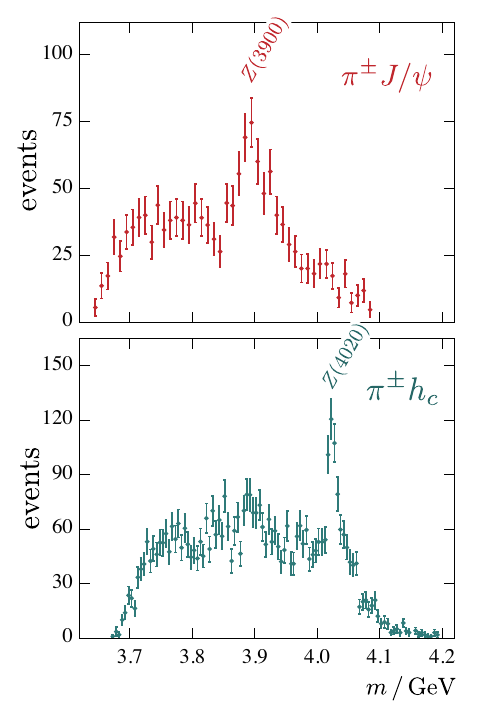}
\captionof{figure}{\label{fig:zc}Number of events collected as a function  of invariant mass of $\pi^\pm J/\psi$ (top)~\cite{Ablikim:2013mio} and $\pi^\pm h_c$ (bottom)~\cite{Ablikim:2013wzq}. In each case one observes a clear narrow enhancement that may be interpreted as a new hadron state.} 
\end{Figure}

These new states can also, in principle, be produced in the weak decay of heavy
mesons containing a bottom quark.  Strangely, recent experimental data yields no evidence 
of $Z(3900)$ production in such decays~\cite{Chilikin:2014bkk}. Instead, signals for still further new
states of higher mass are observed~\cite{Aaij:2014jqa,Chilikin:2013tch,Mizuk:2008me,Chilikin:2014bkk}.  
A related process is the decay of heavy baryons containing a bottom quark,
and here, equally as surprising, we find what appears to be a resonating
proton-$J/\psi$ system~\cite{Aaij:2015tga}.  This hadron is a candidate for a pentaquark.
While the origin of these new states is not yet firmly established, they
present a serious challenge to the simple rules for constructing mesons and
baryons that we previously believed were obeyed by QCD.

The pattern of conventional mesons nicely replicates itself for each flavor
of quark:  many structures that appear in the spectrum of light quarks
(up, down, strange) reappear for charm quarks at the 3 GeV scale, and again
for bottom quarks at the 10 GeV scale.  One might also expect that any spectrum
of hybrids, tetraquarks, or other novel constructions, should have recurrent
patterns for different quark flavors.  In fact bottom-quark analogues of the
charged tetraquark candidates in charmonium have been reported~\cite{Belle:2011aa}.
Historically, these observations preceded those in charmonium.

Like tetraquarks and pentaquarks, another class of hadrons that appear to be
allowed by the fundamental interactions of QCD are quark-gluon hybrids, in which 
gluons and quarks play a role in setting the quantum numbers of the hadron.
A subset of possible hybrid mesons have 
a unique experimental signature: exotic $J^{PC}$ not accessible to a $q\bar{q}$ 
pair.  While there are experimental indications of exotic hybrid 
candidates~\cite{Meyer:2015eta,Ivanov:2001rv,Alekseev:2009xt,Kuhn:2004en, Adams:2011sq}, 
no firmly-established spectrum of hybrid mesons has been discovered.

In parallel to the experimental work discussed above, theoretical efforts
are underway to understand whether QCD \emph{predicts} the existence of
hadrons which go beyond the quark-antiquark meson and three-quark baryon 
rule, or whether the collective behavior of quarks and gluons excludes the 
construction of more exotic combinations. It is to such
calculations that our attention now turns.

\section*{Rules derived from Quantum Chromodynamics}

Much of our understanding of hadrons is informed by models, which
may be motivated by features of QCD, by empirical observations, or both.
A goal is to develop an understanding that is based on rigorous
calculations of the interaction of quarks and gluons through the
equations of QCD.  However, the strongly-coupled nature of QCD makes 
techniques that are practical for calculating weak and electromagnetic
interactions ineffective for predicting properties of hadrons that
emerge from QCD.  We need a different approach, one that utilizes
the fact that all fundamental particles, including quarks and gluons
in QCD, are more correctly thought of as fluctuating quantum fields. The quantum aspect of
the theory is embodied in the fact that observable consequences follow from a
sum over all possible configurations in space and time that these fields can take. The method known as {\it lattice QCD} makes the approximation of considering these fields on a discrete
grid of points describing a restricted region of space-time.  A finite, but
large, number of possible configurations of the fields on this grid can
be generated using random sampling on a computer, and a good
approximation for observable hadron properties obtained.  The volume
of the grid and number of field configurations required to 
achieve useful precision demands significant computational resources.
Total computational times of several teraflop-years are not unusual for 
contemporary calculations, with such efforts making use of leadership 
supercomputing facilities -- future precision lattice QCD calculations of 
increased sophistication will require petaflop-scale machines.

Lattice QCD has been applied with substantial success to a broad range
of processes involving hadrons \cite{Kronfeld:2012uk}, including the spectrum and internal structure of the lightest hadrons
\cite{Kronfeld:2012ym}, the behavior of hadrons at non-zero temperature, relevant
in collisions of heavy ions \cite{Ding:2015ona}, and heavy flavor decays, in which a heavy
quark confined inside a meson decays through the weak interaction \cite{Aoki:2013ldr}.

\subsection*{Lattice QCD as a tool for hadron spectroscopy}
\label{sec:hybrid_rule}

\begin{figure*}
\centering
\includegraphics[width=1\linewidth]{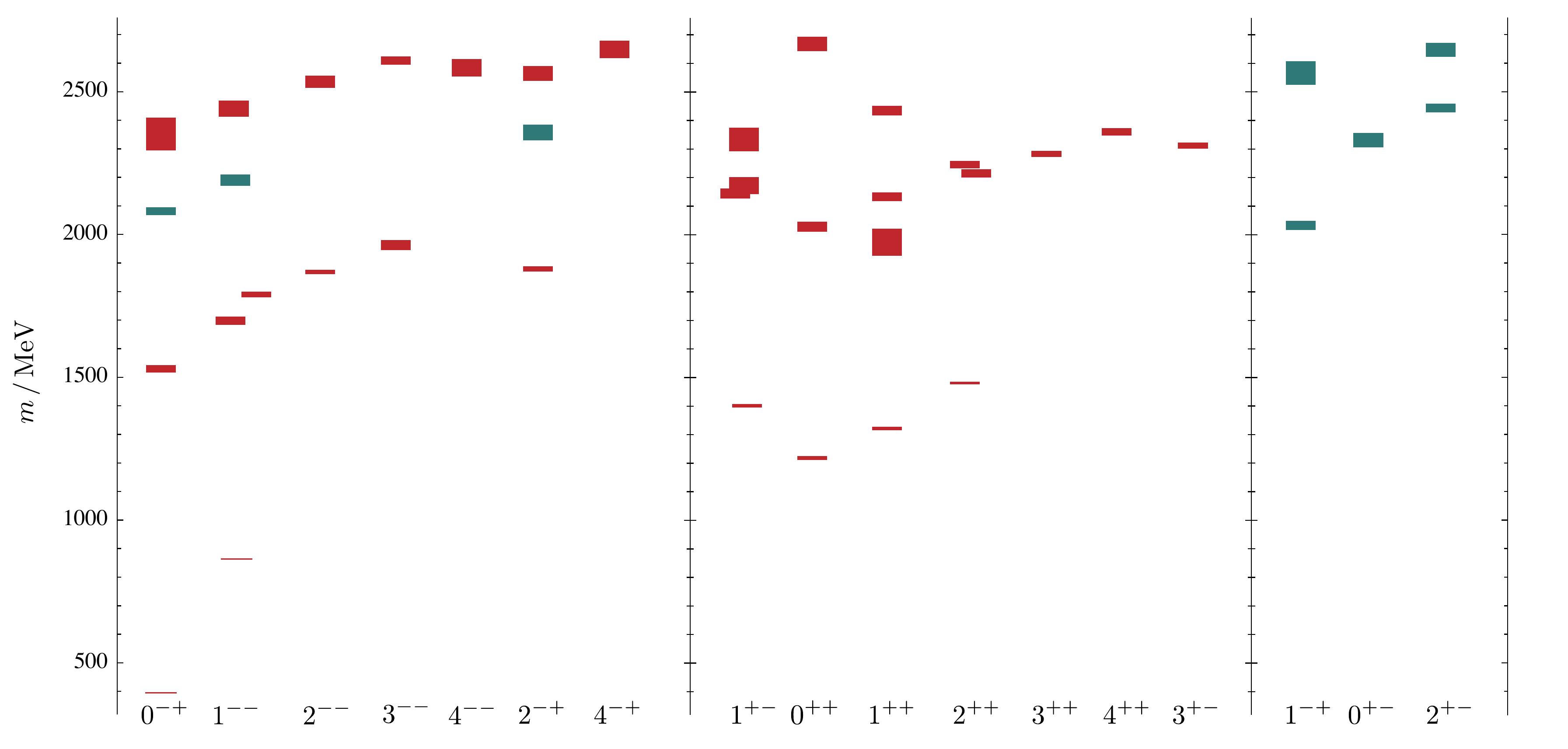}
\caption{\label{fig:lattice_spec}Spectrum of meson states computed using {\it lattice QCD} with light quark masses such that $m_\pi = 392 \,\mathrm{MeV}$~\cite{Dudek:2013yja}. The spectrum features sets of states compatible with the $nL$ assignments of a $q\bar{q}$ model (see Figure~\ref{fig:charmonium}), but also (shown in blue) states which do not have a place in such a model. These states have an interpretation as {\it hybrid mesons} in which a $q\bar{q}$ pair is partnered with an excitation of the gluon field~\cite{Dudek:2011bn} -- their presence suggests a new rule of hadron construction which includes gluons.  (The height of each box represents the estimated uncertainty in the calculation.)} 
\end{figure*}

Our interest is in the determination of properties of \emph{excited} hadrons, 
where obtaining a high degree of numerical precision is an issue that
is secondary to the more basic question of whether certain states
exist or do not. In the past few years we have seen significant progress in overcoming the novel set of challenges posed by these calculations. Exploration of the excited hadron spectrum is possible using an
approach in which each state in the spectrum is produced by
a different combination of quark and gluon field constructions, and in
order for this method to be successful, a large set of possible 
constructions is required. The dynamics of QCD, implemented by the sum over possible field configurations, determines which combination of constructions is present in each state in the spectrum. A scheme outlined in Ref.~\cite{Dudek:2009qf, Dudek:2010wm} 
includes many constructions resembling $q\bar{q}$ pairs with various 
orbital motions and radial wavefunctions, motivated by the success of the $q\bar{q}$ ``rule" in describing the experimental hadron spectrum. More elaborate structures are possible, though, and Ref.~\cite{Dudek:2009qf, Dudek:2010wm} included several that feature the gluon field in a non-trivial way, inspired by the possibility that hybrid mesons may be allowed by QCD.

This large set of constructions, coupled with advances in computational techniques \cite{Peardon:2009gh},
and application of state-of-the-art computing hardware \cite{Egri:2006zm, Clark:2009wm}, 
led to the pioneering results presented in Fig.~\ref{fig:lattice_spec} for the spectrum of mesons 
constructed from light up and down quarks.
The computational challenges of these calculations currently require
the utilization of masses for the lightest quarks that are heavier than the physical 
up and down quark masses, which leads to a systematic shift in computed meson masses.  However, since the 
immediate goal is to understand the underlying QCD dynamics by studying the 
pattern of states, rather than precisely predicting the mass of each meson, the computed 
spectrum allows us to develop intuitive rules for constructing hadrons that
generally apply for quarks of any mass.

The spectrum presented in Fig.~\ref{fig:lattice_spec} qualitatively reproduces many of
the features of the experimental light meson spectrum, and further it
reflects the simple picture of quark-antiquark mesons, with the bulk of the
states fitting into the pattern expected for states excited with
increasing amounts of orbital angular momentum and/or excitations
in the radial quantum number. There are some notable exceptions to this pattern however, in particular the
$0^{-+}$, $1^{--}$ and $2^{-+}$ states between 2.1~GeV and 2.4~GeV do not obviously have an explanation, and most strikingly there is a clear spectrum of states with exotic $J^{PC} = 1^{-+}$, $0^{+-}$ and $2^{+-}$, which
cannot be constructed from a $q\bar{q}$ pair alone. 

These additional mesons, which go beyond the set predicted by the
$q\bar{q}$ rule, have a natural explanation as quark-gluon hybrid mesons.
Previously, estimates for the spectrum of hybrid mesons came only from
models, which made educated guesses for the behavior of the strongly
coupled gluons inside a hadron -- different guesses led to very different
predictions for the number and mass of hybrid states \cite{Horn:1977rq,
Isgur:1984bm, Jaffe:1985qp, Barnes:1982tx, Chanowitz:1982qj,
General:2006ed, Guo:2008yz}.  Using the lattice QCD technique we are now
able to predict a definitive pattern of states, directly from the
fundamental interactions as prescribed by QCD.  Further calculations
\cite{Liu:2012ze, Moir:2013ub, Dudek:2011tt}, performed with larger values
of the quark mass, up as high as the charm quark mass, show the same
pattern of hybrid mesons, and they are found to be consistently 1.3~GeV
heavier than the lightest $J^{PC}=1^{--}$ meson. The particular pattern of
states and the simple mass gap, leads to a new ``rule" of hadron
construction for hybrid mesons namely:  combine $q\bar{q}$ constructions
with a gluonic field that has $J^{PC}=1^{+-}$ and a mass of about 1.3~GeV
to form the spectrum of hybrid mesons in QCD. This is the first example of
a ``rule'' following from a QCD calculation rather than by being inferred
from experimental observations~\cite{Dudek:2011bn}.

Of course this rule must be verified by producing and studying hybrid
mesons in the laboratory, and many current and near future experiments
include searches for these states in their programs. Some hybrid meson
candidates have already been observed experimentally in both the light
meson sector~\cite{Meyer:2015eta,Ivanov:2001rv,Alekseev:2009xt,Kuhn:2004en,
Adams:2011sq} and in the charm region. For example, the $Y(4260)$ discussed
in the previous section has $J^{PC}=1^{--}$, approximately the right mass
relative to the $J/\psi$, and seems to appear in addition to the expected
$q\bar{q}$ excitations. The new rule of hybrid meson construction would
have this meson partnered with states of $J^{PC}= (0,1,2)^{-+}$ at a
similar mass -- searches for these are underway.

\subsection*{Calculating how hadrons decay}

\begin{SCfigure*}
\centering
\includegraphics[width=1.2\linewidth]{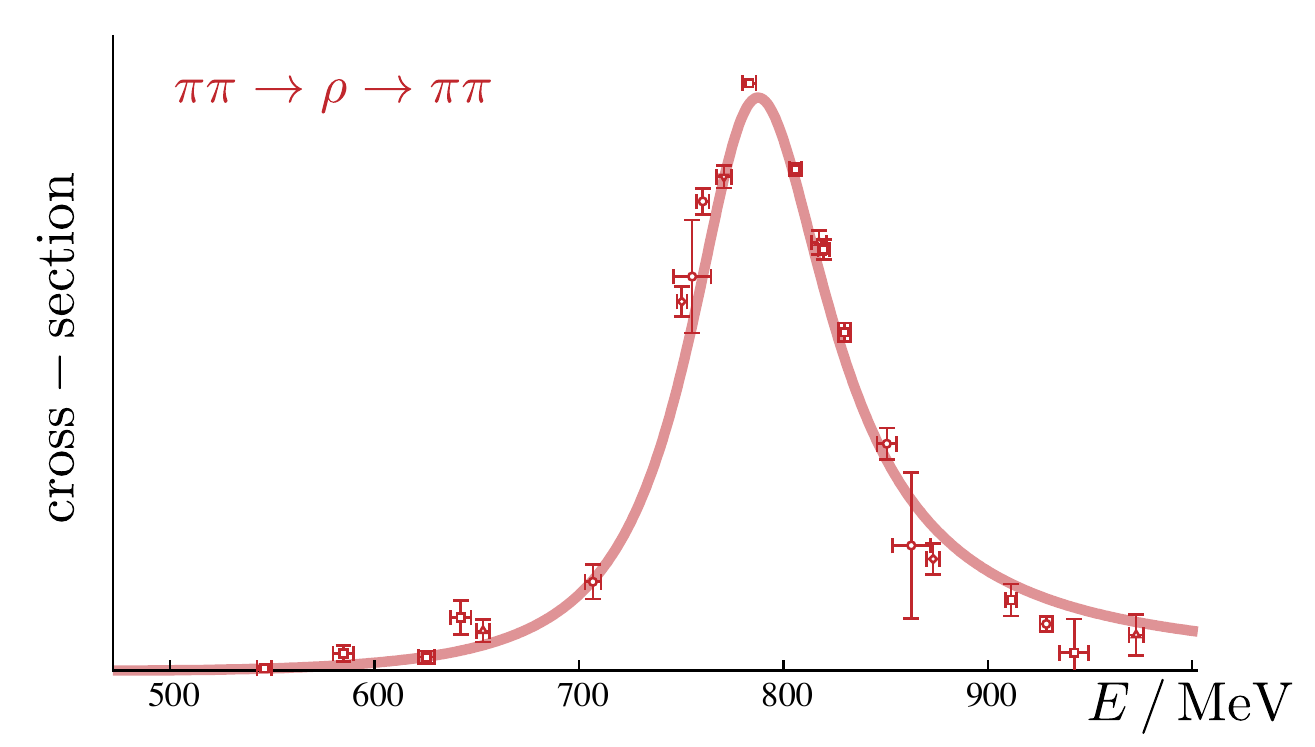}
\caption{\label{fig:xc_pipi}Cross-section (in arbitrary units) for $\pi\pi \to \pi\pi$ with $1^{- -}$ quantum numbers calculated using {\it lattice QCD} with light quark masses such that $m_\pi = 236\,\mathrm{MeV}$ \cite{Wilson:2015dqa}. The $\rho$ resonance is clearly observed as a bump, and from its position and width the mass and decay rate of this excited state can be extracted. } 
\end{SCfigure*}

These calculations of the excited meson spectrum within QCD represent a major step forward
in our understanding of hadron spectroscopy, but they still make approximations which fail
to capture an important feature of excited hadrons -- that they are
{\em resonances}, decaying rapidly to lighter hadrons. As can be seen in
Figs.~\ref{fig:epem_hadrons} and~\ref{fig:zc}, in simple cases, excited
states appear as characteristic bumps in the rate of observation of
certain final state mesons, and lattice QCD calculations should be capable of reproducing this behavior.

Experimentally, resonances are often observed to decay preferentially into
certain sets of mesons and not others, and these patterns can be used to
infer details of the resonant state's internal structure. To be able
to calculate the decay properties of the excited hadrons directly from
QCD would provide a powerful tool for interpreting experimental data. In the
case of predictions of previously unobserved excitations, it may also
provide suggested decay channels to be examined in experimental searches.

Extending the calculations described above to correctly account for the decay
of excited states is possible \cite{Luscher:1990ux, Rummukainen:1995vs, Kim:2005gf, Christ:2005gi, He:2005ey, Hansen:2012tf, Briceno:2012yi, Guo:2012hv} 
but challenging, and serious efforts
have only recently begun \cite{Aoki:2007rd, Feng:2010es, Lang:2011mn, Aoki:2011yj, Pelissier:2012pi, Fu:2011wc, Lang:2012sv, Dudek:2012xn}. As an example of what can be achieved, in
Fig.~\ref{fig:xc_pipi} we present the cross-section for two pions to form the 
lightest $1^{--}$ resonance, known as the $\rho$, and then to decay back 
into two pions. A clear bump is observed whose position and width provide 
the mass and decay rate of the $\rho$.

These rapidly maturing theoretical techniques will be required to study the new charmonium
mesons, discussed earlier, within QCD. The observed enhancements are
seen only in specific final states, which implies that the ability to predict how
hadrons decay directly from QCD will be an essential component in interpreting
experimental data in the quest to develop the rules for constructing hadrons.

\section*{Towards a unified set of rules}

Much of what we know about what emerges from strongly coupled QCD has come
from studying patterns of hadrons organized by mass and quantum numbers
like $J$, $P$, and $C$. These patterns suggest quarks of several flavors
which may be combined with a single antiquark to form mesons -- a rather
simple ``rule'' of hadron construction. The theory of QCD is not limited to
such simple constructions however, and making a definitive statement about
the existence of mesons with four-quark or quark-gluon hybrid structure
will require observing a spectrum of additional mesons that cannot be
explained by the $q\bar{q}$ rule. In particular we will need to observe a
set of states with unusual flavor and/or $J^{PC}$.

\subsection*{Finding a pattern of hadrons is essential}

Contemporary technology has enabled experimental investigations at an
unprecedented level of statistical precision, which provides the capability
to discover more rare and interesting phenomena.  However, this
demands a refined precision with which we attempt to interpret experimental
data.  For example, one needs to be certain that the same logic that allows
one to conclude the presence of conventional charmonium mesons in the total
hadronic cross section, also applies when one is
examining the cross section for a single exclusive process that is two
orders of magnitude smaller (see e.g. Fig 2.).  Such precise experimental data make one
susceptible to effects that can mimic the experimental signature of a 
new hadron, but which in fact have a more prosaic origin~\cite{Bugg:2008wu,Guo:2015umn,Swanson:2014tra,Szczepaniak:2015eza,Wang:2013hga}.
This underscores the importance of experimentally establishing a {\it pattern} of
hadrons:  the interpretation of any single state as a new and exotic construction will certainly be questioned.  However, the experimental observation
of an ordered spectrum of states becomes harder to explain as a
misinterpreted experimental artifact.

Likewise, theoretical efforts in lattice QCD must continue in their
attempts to compute the complete set of possible hadrons allowed by QCD,
and to identify pattens of states within that spectrum. Recent advances
have enabled us to develop a simple rule, as stated in
Section~\ref{sec:hybrid_rule} that describes how QCD constructs hybrid
mesons and baryons, in an extension of what we had already for conventional
mesons and baryons -- this new rule must be verified by observing an
experimental spectrum of hybrids.  Lattice QCD can also be used to
calculate decay properties of hadrons, and identifying particular
characteristic decays of hybrid mesons will guide experimental searches and
aid in interpretations of data.  As has been done with hybrids, lattice QCD
needs to determine if QCD predicts a spectrum of tetraquark and pentaquark
states. A particular priority is in the heavy quark sectors, where, as we
have discussed above, there is recent experimental evidence for such
objects. The ability within lattice QCD to vary arbitrarily the mass of the
quarks, allows us to identify how the ``rules'' of hadron construction
vary, and to identify possible common behaviors between the heavy
charmonium system and the lighter mesons.

\subsection*{A global experimental program}

Establishing a spectrum of hadrons beyond those described by the 
simple $q\bar{q}$ and $qqq$ rules will require the combined
efforts of multiple present and future experiments.  There is a
spectroscopy program within nearly every particle physics collaboration
worldwide.  We list the details of a selection of several past, present, and future
experiments, primarily those whose work is referenced in this article,
in Box 1 as an illustration of the breadth of the worldwide effort.

Most of the recently observed new hadrons are so-far observed in 
only a single production or decay process. Observation of
the same state in multiple production and decay modes almost certainly
rules out a misinterpretation of experimental data due to some
process-dependent phenomenon and solidifies the evidence for a new
hadron.  Therefore the best current routes to explore new states in the
charm sector are by comparing results from $e^+e^-$ collisions (BESIII, Belle, BelleII) 
and production in $B$-meson decay (LHCb, Belle, BelleII).  Supplementing 
these with results from novel production mechanisms, like proton anti-proton 
annihilation ($\overline{\mathrm P}$ANDA), would be extremely valuable.

Experiments aimed at exploring different energy regimes and quark flavors are essential for a
complete understanding of the meson spectrum, as one expects the underlying
patterns of states to be independent of quark mass.  A variety of present and
future experiments will allow access to both the charmonium system (BESIII, Belle, Belle II, LHCb),
and the analogous system of bottom quarks, bottomonium (Belle, BelleII).
Mesons constructed from light quarks can be produced in
decays of heavier mesons and therefore can be studied at all of the
previously-mentioned facilities; they can also be produced at experiments dedicated 
to the study of lighter systems (COMPASS, GlueX). Discovery of light quark hybrids 
would suggest the existence of heavy quark hybrids and further motivate 
dedicated searches for these states.

With continued coordinated experimental and theoretical investigations
we hope to define a complete set of rules for building hadrons that both 
describes what is observed in nature and can be derived directly
from QCD. In doing so we aim to understand how what seems to be
a simple spectrum of hadrons emerges from the complex interactions of 
quarks and gluons in QCD.

\paragraph{Acknowledgements:}  JJD acknowledges support provided by U.S. Department of Energy contract DE-AC05-06OR23177 under which Jefferson Science Associates (JSA) manages Jefferson Lab and the Early Career award contract DE-SC0006765.  MRS and REM are supported by U.S. Department of Energy contract DE-FG02-05ER41374.  MRS acknowledges the JSA Sabbatical Leave Support Program. The authors thank L.~Weinstein for useful comments on the initial draft of this manuscript.  All authors contributed equally to this manuscript.  Correspondence should be directed to MRS at {\it mashephe@indiana.edu}.  The authors declare no competing financial interests.

\end{multicols}

\begin{center}
\subsection*{Box 1}
\fbox{\begin{minipage}{1.0\linewidth}
Information about a selection of experiments and their hadron spectroscopy programs, which 
typically represent only a fraction of each collaboration's research efforts. 

\begin{itemize}

\item {\bf BaBar} (Menlo Park, CA, USA):  $e^+e^-$ collisions at bottomonium energies; 
discoveries of the $Y(4260)$ and $Y(4360)$; finished collecting data in 2008.

\item {\bf Belle} (Tsukuba, Japan):  $e^+e^-$ collisions at bottomonium energies; 
discovery of the $X(3872)$, $Z(3900)$, $Z(4430)$, and $Z_b$ states; finished collecting data in 2010.

\item {\bf Belle II} (Tsukuba, Japan): an upcoming continuation of the Belle experiment that
will provide much higher intensity $e^+e^-$ collisions than achieved at Belle.

\item {\bf BESIII} (Beijing, China):  $e^+e^-$ collisions at charmonium energies;  
direct production of the $Y(4260)$; discovery of the $Z(3900)$ and $Z(4020)$; ongoing.

\item {\bf COMPASS} (Geneva, Switzerland):  high-intensity meson beams on nuclear targets; 
searches for unusual light-quark mesons; discovery of the $a_1(1420)$; ongoing.

\item {\bf GlueX} (Newport News, VA, USA):  polarized photon beam on a nuclear target; 
searches for light-quark hybrid mesons; data collection is beginning now.

\item {\bf LHCb} (Geneva, Switzerland):  high-energy, high-intensity proton-proton collisions, 
specializing in $B$-meson decays; measurement of resonant nature of the $Z(4430)$; 
discovery of pentaquark candidates; ongoing.

\item {\bf \boldmath $\overline{\mathrm P}$ANDA} (Darmstadt, Germany): proton-antiproton 
collisions at charmonium energies; exploration of charmonium and light-quark mesons; upcoming.

\end{itemize}

\end{minipage}}
\end{center}

\end{document}